\documentstyle[12pt,twoside,fleqn,espcrc1,epsfig]{article}

\newcommand{\AmS}{{\protect\the\textfont2
  A\kern-.1667em\lower.5ex\hbox{M}\kern-.125emS}}

\title{Baryonic contributions to $e^+e^-$ yields in a hydrodynamic model
       of Pb+Au collisions at the SPS}

\author{\underline{P.\ Huovinen}\address{Department of Physics,
                                      University of Jyv\"askyl\"a, Finland}%
        and
        M.\ Prakash\address{Department of Physics \& Astronomy, SUNY at
                            Stony Brook, Stony Brook, USA}}

\begin{document}
% typeset front matter
\maketitle

\begin{abstract}
We analyze $e^+e^-$ yields from matter containing baryons in addition
to mesons using a hydrodynamic approach to describe Pb+Au collisions
at 158 $A$ GeV/c. We use two distinctly different
$e^+e^-$ production rates
to provide contrast. Although the
presence of baryons leads to significant enhancement of $e^+e^-$
emission relative to that from mesons-only matter, the calculated
results fall below the data in the range $400 < M_{e^+e^-}/{\rm MeV} < 600$.
The calculated results are, however, only 1.3--1.5 standard deviations
below the data, which may not be statistically significant.
\end{abstract}

\section{Theoretical Approach}
The observation of low-mass dielectron excess over conventional sources
by the CERES collaboration~\cite{Lenkeit99,Ravinovich98} has spurred
considerable theoretical activity. In addition to the microscopic
production rates, confrontation of theory with data
requires a description of the
space-time evolution of the produced matter.
We have used a one-fluid hydrodynamical description, which
is constrained to reproduce the measured hadron spectra~\cite{Huovinen98a}.
The initial state of the hydrodynamic evolution,
which is assumed to be sufficiently thermalized,
 is parametrized to reproduce
both the baryonic and mesonic components.
The adiabatic expansion is then governed by the laws of hydrodynamics
(including baryon number and strangeness conservation)
and the input
equation of state (EOS). In the calculations
reported here, the EOS admits
a phase
transition to the quark-gluon plasma at a critical temperature
$T_c = 200$ MeV. The
hadronic part of the EOS includes particles and resonances up to 2 GeV.
The system is assumed to maintain both local thermal and
chemical equilibrium until freeze-out. The freeze-out criterion employed,
energy density
$\epsilon_f = 0.15$ GeV/fm$^3$, corresponds to an average
freeze-out temperature of $T_f=140$ MeV.
Admissible changes in both $T_c$ and $T_f$ do not affect our
conclusions~(see~\cite{Huovinen98b}).

\begin{figure}
\begin{center}
  \begin{minipage}{7.8cm}
        \epsfxsize 7.1cm \epsfbox{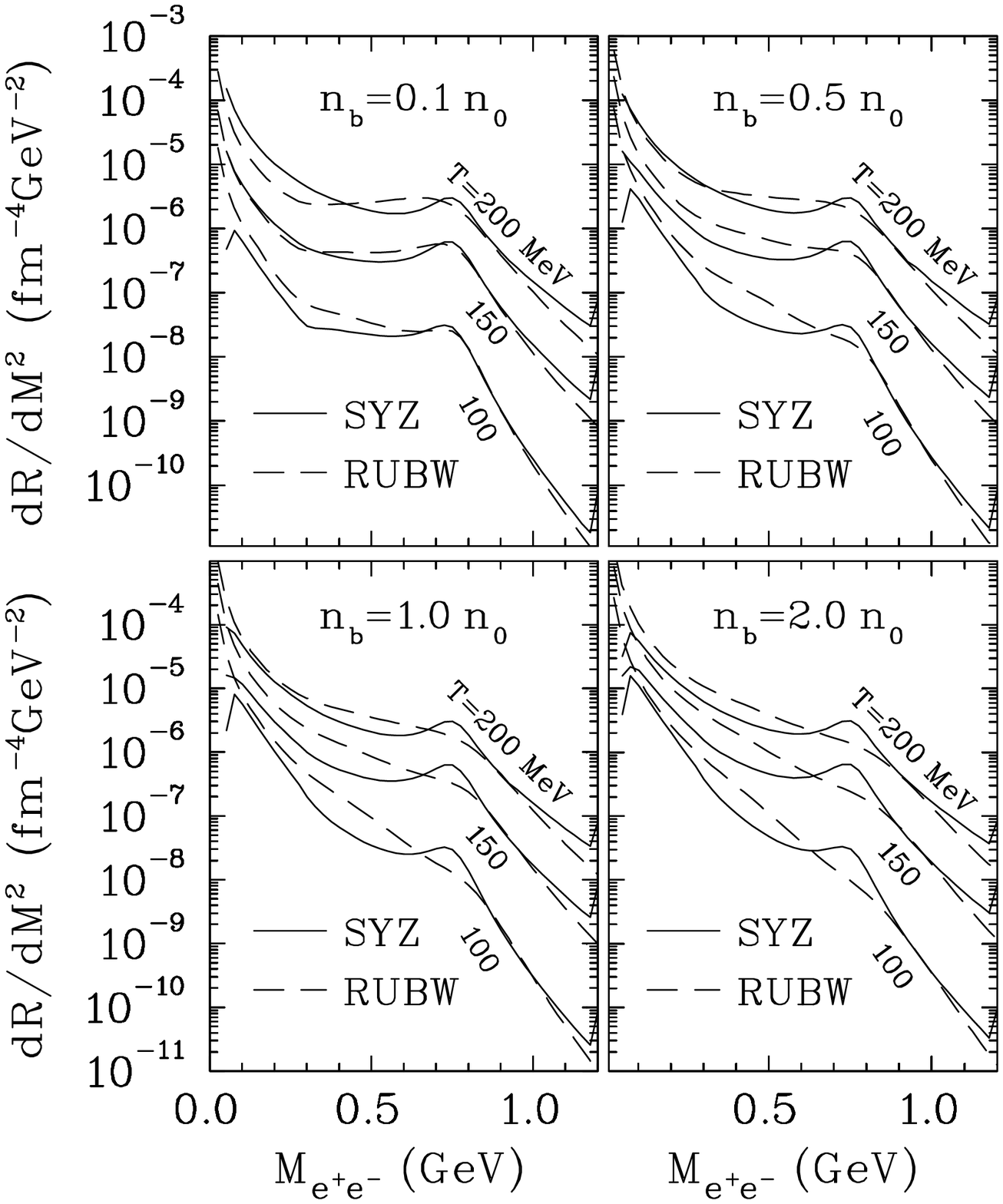}
        \hfill
  \caption{The $e^+e^-$ production rates at different temperatures and
                baryon densities versus pair invariant dielectron mass.
		The solid (dashed)
                lines are results from Steele, Yamagishi, and Zahed (SYZ)
                (Rapp, Urban, Buballa, and Wambach (RUBW)).}
  \label{rates}
  \end{minipage}
   \hfill
  \begin{minipage}{7.8cm}
        \epsfxsize 7.5cm \epsfbox{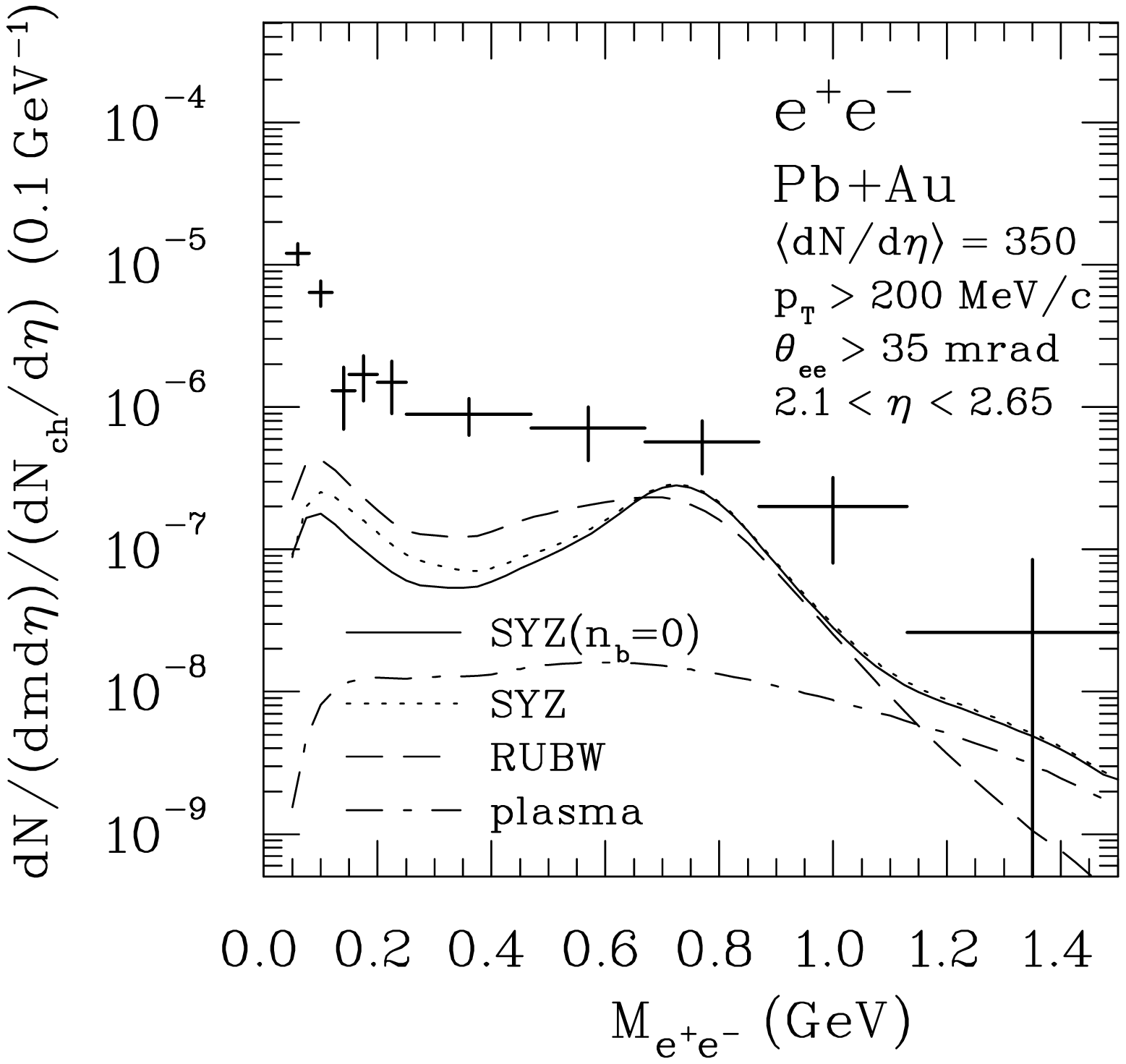}
        \hfill
   \caption{Dielectron yields from the hadronic part and plasma part
              of the fireball. Solid lines
              are results for matter without baryons with
              the SYZ rates. Also shown are results in matter with baryons
              (short-dashed lines: SYZ rates; long-dashed lines:
              RUBW rates).
              Kinematic cuts and detector resolution are incorporated.
              The data shown are preliminary~\protect\cite{Lenkeit98}.}
  \label{thermal}
  \end{minipage}
\end{center}
\end{figure}

We  use $e^+e^-$ production rates from the calculations of Steele,
Yamagishi and Zahed (hereafter SYZ~\cite{Steele97})and Rapp, Urban, Buballa and
Wambach (hereafter RUBW~\cite{Rapp98}). SYZ use experimentally extracted
spectral functions and on-shell chiral reduction formulas coupled with a virial
expansion scheme. For baryon number density $n_b = 0$,  these rates reproduce
those of Gale and Lichard~\cite{Gale94}. The RUBW rates are based on a
many-body approach, in which phenomenological interactions are used to
calculate the $\rho$-meson spectral function in matter containing baryons.
These two rates represent contrast both in terms of their input physics and the
theoretical techniques employed. They also differ significantly in their
absolute magnitudes (see Fig.~\ref{rates}). The distinguishing features of the
SYZ rates are: (1) Enhancements relative to the baryon-free case are of order
2-3 and are restricted to $M_{e^+e^-}/{\rm MeV} < 500$. (2) The prominent
signature at the $\rho$-meson vacuum mass persists at nearly all values of $T$
and $n_b$. In contrast to SYZ, the two most striking features of the RUBW
results are: (1) Rates are significantly larger than those of SYZ in the range
$ 200 < M_{e^+e^-}/{\rm MeV} < 600$. (2) The tell-tale signature at the
$\rho$-meson vacuum mass is weakened, predominantly with increasing $n_b$.
\section{Results}
In Fig.~\ref{thermal}, we show the results of dielectrons radiated during the
lifetime of the fireball. These results are folded with the cuts
and resolution of CERES.
Our calculation is tuned to reproduce the hadronic results of NA49
and yields an average multiplicity of
$\langle dN_{ch}/d\eta \rangle \cong 330$ within the CERES acceptance
region. The CERES collaboration finds both the shape of the spectrum
and the yield scaled with multiplicity to vary with
multiplicity~\cite{Ravinovich98}. We have therefore opted to compare
our results with the preliminary data~\cite{Lenkeit98} from nearly central
collisions with $\langle dN_{ch}/d\eta \rangle = 350$.

The contribution from the quark-gluon
plasma is about an order of magnitude below the
hadronic contributions. The SYZ rates with
baryons are about a factor of $2$ larger than those without baryons,
but mostly below $M_{e^+e^-}=400$ MeV. This translates to an
enhancement of about a factor of two or less relative
to the baryon-free case. The larger rates of RUBW result in
enhancements of the thermal yield, even up to $M_{e^+e^-}=300-600$
MeV, by a factor of about three relative to those in mesons-only
matter.

In addition to thermal pairs, the measured yield contains contributions
from meson decays after freeze-out. This background was calculated
from the distributions and abundances of hadrons at freeze-out given
by our calculations. The only exception is the $\phi$-yield, which is
suppressed by a factor 0.6 to achieve consistency with the
data~\cite{Friese97}. The resulting background is in
agreement with the background estimated by the CERES collaboration.

\begin{figure}
\begin{center}
  \begin{minipage}[t]{7.8cm}
        \epsfxsize 7.5cm \epsfbox{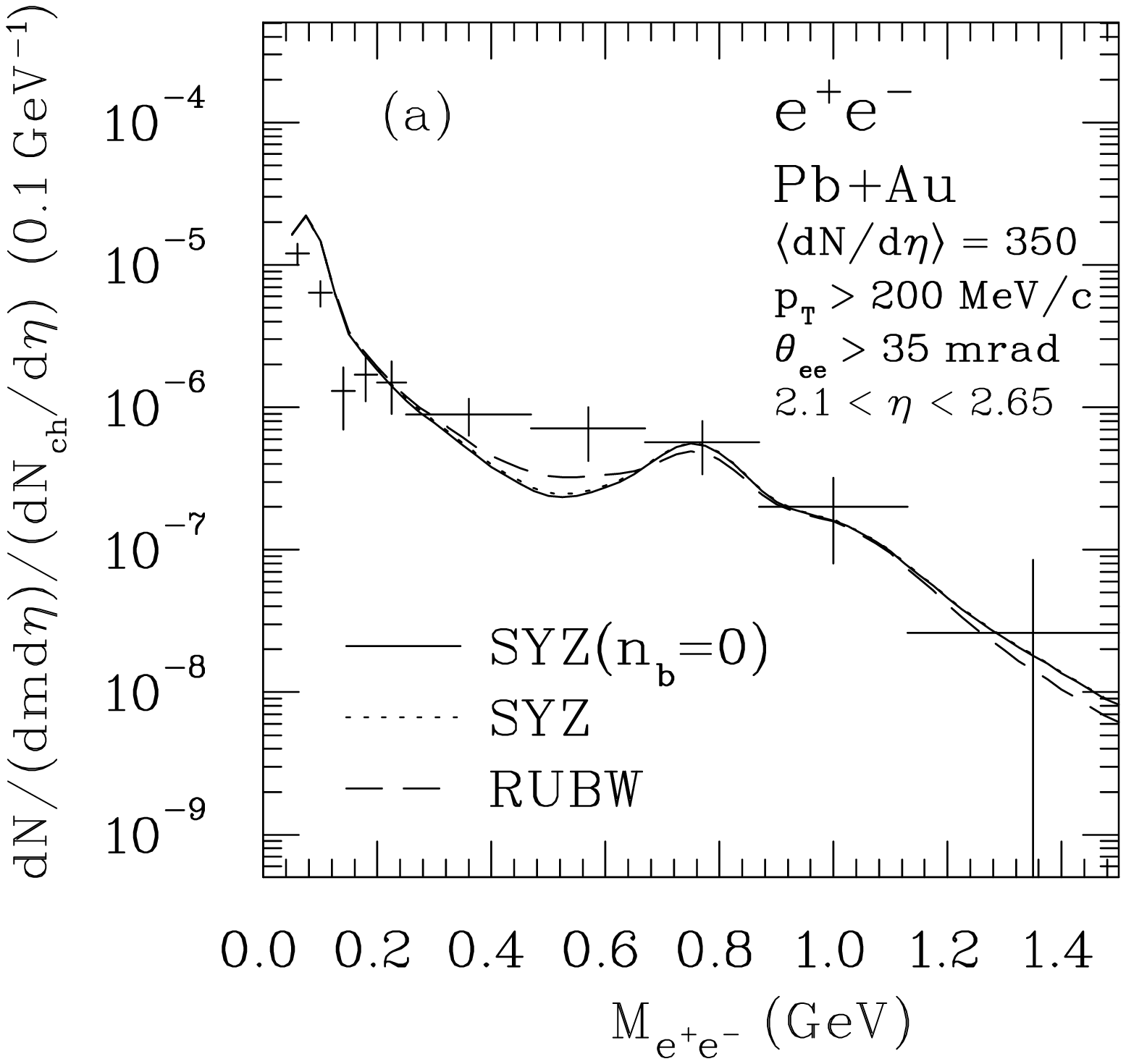}
        \hfill
%  \caption{Calculated total dielectron mass spectra
%              compared with
%              preliminary data~\protect\cite{Lenkeit98}.
%              Solid lines are results for matter without baryons and are
%              for the rates of SYZ, which agree with those
%              of GL. Results in matter with baryons (short-dashed lines
%              from the SYZ rates and long-dashed lines from those of
%              RUBW) are also shown. Kinematic cuts and detector
%              resolution are incorporated.}
%  \label{result}
  \end{minipage}
   \hfill
  \begin{minipage}[t]{7.8cm}
        \epsfxsize 7.5cm \epsfbox{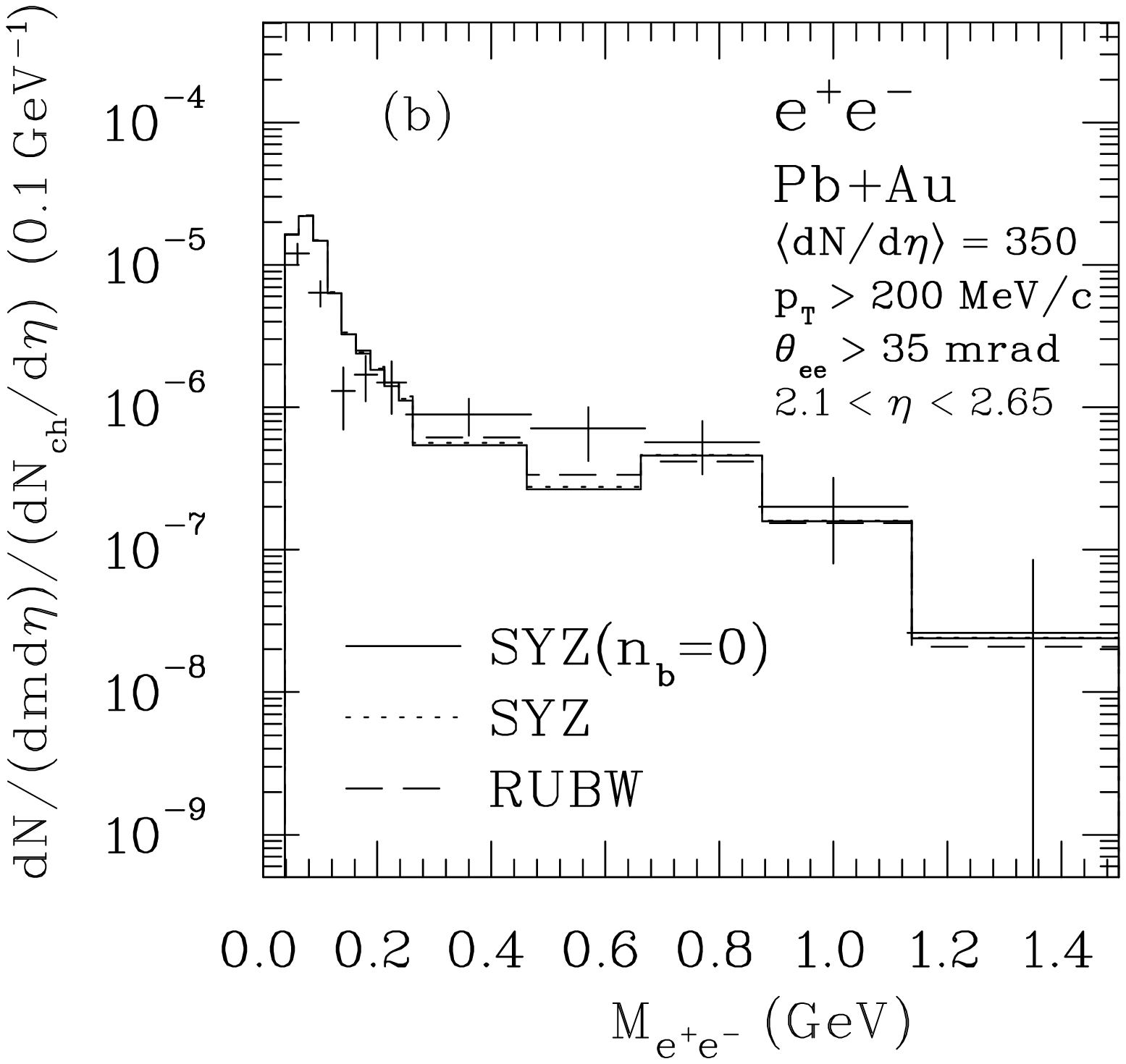}
        \hfill
%   \caption{As Fig.~\ref{result}, but the calculated result is presented in
%            in larger bins to make the
%            comparison to the CERES data more accurate.}
%  \label{results2}
  \end{minipage}
  \caption{(a) Calculated total dielectron mass spectra
              compared with
              preliminary data~\protect\cite{Lenkeit98}.
              Solid lines are results for matter without baryons with
              the SYZ rates. Also shown are results in matter with
              baryons (short-dashed lines:  SYZ rates;  long-dashed
              lines: RUBW rates) . Kinematic cuts
              and detector resolution are incorporated.
             (b) Like (a), but the calculated results employ
              bins to contrast with those used by CERES.}
 \label{result}
\end{center}
\end{figure}

The total yield (sum of the thermal and background
contributions) is presented Fig. \ref{result}(a). Results for
the SYZ rates with and without baryons are virtually indistinguishable
from each other, despite significant enhancements in the
microscopic rates
for $M_{e^+e^-} < 400$ MeV. In this mass region, the
Dalitz decay backgrounds are an order of magnitude larger than the
thermal yields and entirely mask the baryonic contributions.
The RUBW microscopic rates, being larger than those of SYZ in
the region below the $\rho$-mass, lead to total yields that are
somewhat distinguishable from the case without baryons, but lie below the
data roughly by a factor of two.

It is instructive to bin the calculated results
to contrast with the bins used by the
CERES collaboration (see Fig.~\ref{result}(b)).
The calculated spectra fall below the data at only two points. At
$M = 360$ MeV, the results are close to the experimental lower limit.
At $M= 570$ MeV, the difference between the data and the
calculated result is only about 1.5 standard deviations for the SYZ rates
with and
without baryonic contributions, whereas the use of the RUBW rates
leads to an yield which is 1.3 standard deviations below the data. Given the
uncertainties, these
differences may not be statistically significant.
We thus conclude that
thermal production of electron pairs may well  be  large enough to
account for the observed enhancement.

\section{Summary}
We have calculated $e^+e^-$ emission in Pb+Au collisions at 158 AGeV/c
using two different dielectron production rates within the framework
of hydrodynamics. The rates calculated by SYZ include baryonic
contributions arising from pion-nucleon interactions and those of RUBW
account for additional in-medium modifications, which leads to a
substantial broadening of the $\rho$-meson spectral function. We found
that the additional contributions due to baryons in the rates of SYZ
give modest contributions, but mainly at low values of invariant mass
where the spectrum is dominated by background decays. The final
dielectron spectra with and without baryonic contributions are thus
almost identical. On the other hand, the larger $\rho$-width in the
rates of RUBW leads to comparatively larger yields in the 300--600 MeV
mass region. In all cases, the calculated results are below the data,
but the differences are not large to indicate statistical
significance. The  yield of thermal dielectrons appears to be
large enough to explain the preliminary data.\\

We thank A.~Drees and I.~Tserruya for helping us to put the
CERES data in perspective.

\end{document}